\documentclass[fleqn,usenatbib]{mnras}

\usepackage{newtxtext,newtxmath}

\usepackage[T1]{fontenc}
\usepackage{ae,aecompl}

\usepackage{graphicx}	
\usepackage{amsmath}	
\usepackage{amssymb}	

\newcommand{\ujy}{$\mu$Jy\,}

\title[The H$\alpha$ SFR - RC luminosity relation at $1.4 < z < 2.6$]{The MOSDEF Survey: Calibrating the relationship between H$\alpha$ star-formation rate and radio continuum luminosity at $1.4 < z < 2.6$}

\author[Kenneth J. Duncan et al.]{Kenneth J. Duncan,$^{1,2}$\thanks{E-mail: kdun@roe.ac.uk (KJD)} 
Irene Shivaei,$^{3}$\thanks{Hubble fellow; Email: ishivaei@arizona.edu (IS)} 
Alice E. Shapley,$^{4}$ 
Naveen A. Reddy,$^{5}$ \and
Bahram Mobasher,$^{5}$
Alison L. Coil,$^{6}$  
Mariska Kriek,$^{7}$ and
Brian Siana$^{5}$
\\
$^{1}$ Leiden Observatory, Leiden University, PO Box 9513, NL-2300 RA Leiden, The Netherlands \\
$^{2}$ SUPA, Institute for Astronomy, Royal Observatory, Blackford Hill, Edinburgh, EH9 3HJ, UK \\
$^{3}$ Steward Observatory, University of Arizona, Tucson, AZ 85721, USA\\
$^{4}$ Department of Physics \& Astronomy, University of California, Los Angeles, CA 90095, USA\\
$^{5}$ Department of Physics \& Astronomy, University of California, Riverside, CA 92521, USA\\
$^{6}$ Center for Astrophysics and Space Sciences, Department of Physics, University of California, San Diego, 9500 Gilman Drive., La Jolla, CA 92093, USA\\
$^{7}$ Astronomy Department, University of California at Berkeley, Berkeley, CA 94720, USA}

\date{Accepted XXX. Received YYY; in original form \today}

\pubyear{2019}

\begin{document}
\label{firstpage}
\pagerange{\pageref{firstpage}--\pageref{lastpage}}
\maketitle

\begin{abstract}
The observed empirical relation between the star-formation rates (SFR) of low-redshift galaxies and their radio continuum luminosity offers a potential means of measuring SFR in high redshift galaxies that is unaffected by dust obscuration.
In this study, we make the first test for redshift evolution in the SFR-radio continuum relation at high redshift using dust-corrected H$\alpha$ SFR.
Our sample consists of 178 galaxies from the MOSFIRE Deep Evolution Field (MOSDEF) Survey  at $1.4 < z < 2.6$ with rest-frame optical spectroscopy and deep 1.5 GHz radio continuum observations from the Karl G. Jansky Very Large Array (VLA) GOODS North field.
Using a stacking analysis we compare the observed radio continuum luminosities with those predicted from the dust-corrected H$\alpha$ SFR assuming a range of $z\sim0$ relations.
We find no evidence for a systematic evolution with redshift, when stacking the radio continuum as a function of dust-corrected H$\alpha$ SFR and when stacking both optical spectroscopy and radio continuum as a function of stellar mass.
We conclude that locally calibrated relations between SFR and radio continuum luminosity remain valid out to $z\sim 2$.
\end{abstract}

\begin{keywords}
galaxies: evolution -- galaxies: star formation -- radio continuum: galaxies
\end{keywords}



\section{Introduction}
The total star-formation rate (SFR) within a galaxy is one of the fundamental observable properties required to trace the mass assembly, growth and formation of galaxies through cosmic time.
Unaffected by dust-obscuration, the radio continuum (RC) luminosity is known to strongly correlate with the total SFR of star-forming galaxies \citep{1992ARA&A..30..575C,1998ApJ...507..155C}.
Studies of numerous samples within the local Universe \citep{1992ARA&A..30..575C,1998ApJ...507..155C,Bell:2003bj,2009ApJ...703.1672K,2011ApJ...737...67M,2015A&A...579A.102B} and at multiple radio frequencies \citep{Brown:2017uu,2018MNRAS.475.3010G} have since confirmed this trend, albeit with small variations in the shape of the observed correlation and its intrinsic scatter.

Thanks to a new generation of sensitive and efficient radio telescopes such as the Low Frequency Array \citep[LOFAR;][]{vanHaarlem:2013gi} and MeerKAT \citep{Booth:2009wx}, deep radio continuum surveys are beginning to probe radio luminosities equivalent to deep far-infrared (FIR) surveys \citep[e.g.][]{2010iska.meetE..50R,2017arXiv170901901J} but with significantly better resolution (and hence fewer problems with source confusion).

To exploit the forthcoming deep RC observations as a probe of unobscured SFR out to the highest redshifts, it is essential to ensure that the empirical SFR-radio correlations observed in the local Universe are still valid at the redshifts of interest -- or if they do evolve, accurately measure and understand these effects.
Extending the use of RC emission as a SFR tracer to the early Universe has to-date relied upon the calibration of the RC luminosity to SFR via the empirical correlation between total infrared luminosity (and its relation to total SFR) and RC luminosity; the so-called `far-infrared radio correlation' \citep[FIRRC,][]{1992ARA&A..30..575C}.
Using deep FIR and sub-mm surveys \citep[e.g.][]{Oliver:2012ds}, the FIRRC has been well studied out to $z\sim3$ in a number of different radio and FIR datasets.
Many of these such studies have found evidence for evolution in the \emph{apparent} FIRRC \citep{2010A&A...518L..31I,  2012ApJ...761..140C, 2015A&A...573A..45M, 2017A&A...602A...4D, 2017MNRAS.469.3468C,2018MNRAS.475..827M}, with studies typically finding an excess of radio continuum emission at higher redshifts.
However, the very modest evolution observed combined with the large intrinsic scatter ($>0.3\,\rm{dex}$) in the relation has led many to favour a scenario in which there is no evolution in the \emph{intrinsic} relation \citep{2010ApJ...714L.190S, 2011MNRAS.410.1155B, 2010MNRAS.409...92J, 2014MNRAS.445.2232S}, i.e. that the apparent trends observed in other studies are a result of potential sample population or observational biases. 
In a recent study at lower redshift, \citet{2018MNRAS.480.5625R} have found evidence for systematic variations in the FIRRC relation with respect to stellar mass and specific SFR (sSFR) within the observed population (with radio emission increasing with sSFR). 
However, it is not clear what the fundamental physical causes behind these variations are.
Equally, it is also not yet clear whether such systematic variations would give rise to the observed redshift trends when accounting for the selection effects of high-redshift studies in flux-limited surveys.
For a detailed discussion on the physical processes driving the FIRRC and how those may effect its redshift evolution, we refer the reader to \citet{2010ApJ...717....1L} and \citet{2010ApJ...717..196L}.

The broad range of literature conclusions regarding the shape of the FIRRC and its evolution can be attributed to the strong observational biases present in measurements of the FIRRC at all redshifts.
Studies of the nearby Universe span a limited range of luminosities (both RC and FIR) due to the limited cosmic volume being sampled.
Conversely, studies of the higher redshift population, where we select only the most luminous sources, are affected by Malmquist bias inherent to flux limited samples.
Furthermore, the large beam size of \emph{Herschel} FIR observations means that the deep field observations used to measure the total SFR(IR) at higher redshifts \citep[e.g.][]{Oliver:2012ds} are significantly confusion limited.
This source confusion presents challenges in reliably measuring the FIR flux of these faint high-redshift sources, with significant selection biases that must be taken into account \citep[see e.g.][]{2010ApJ...714L.190S,2017A&A...602A...4D}.
Additionally, at $z > 1$ the robust separation of RC emission from star-formation and AGN from photometry alone becomes increasingly difficult - a result of both degeneracies in spectral energy distribution (SED) fitting \citep{2017MNRAS.469.3468C}, and the still poorly understood link between AGN jet activity and other signatures of accretion \citep{2012MNRAS.421.1569B}.

The observed redshift evolution in the FIRRC may therefore be driven by increased AGN activity (as traced by the RC luminosity) within the sample.
When investigating the evolution of FIRRC out to $z\sim1.5$ as a function of morphology, \citet{2018MNRAS.475..827M} find that the overall observed redshift evolution may in fact be driven by a stronger evolution of bulge-dominated star-forming galaxies (SFG) - with little to no evolution for disk-dominated SFG morphologies.
The increased evolution for bulge-dominated galaxies might be attributed to residual AGN activity in these systems that is not identified by other AGN selection criteria - a scenario consistent with the build up of the black hole-bulge mass correlation over this epoch \citep{2012ApJ...753L..30M}.
As shown by \citet{2019A&A...622A..17S}, the black holes residing in massive galaxies ($>10^{11} \textrm{M}_{\odot}$) in the local Universe are always `switched on' at some level -- such trends may also hold true for equally massive galaxies at higher redshift.

Given the aforementioned uncertainties in the calibration of RC luminosity as a SFR indicator using the FIRRC, direct calibrations through reliable alternative SFR indicators are required if we want to use RC as means of tracing obscuration free SFRs (and understand the causes of the observed FIRRC trends). 
In the local Universe, dust-corrected H$\alpha$ luminosity remains one of the prime SFR indicators \citep{2012ARA&A..50..531K}
and has been demonstrated to provide consistent SFR estimates out to $z \sim 2.6$ \citep{Erb:2006ke,2010ApJ...712.1070R,2015ApJ...815...98S,2016ApJ...820L..23S}.
In this paper we attempt the first such calibration at the peak of the cosmic star-formation rate history ($z\sim2$).

The MOSFIRE Deep Evolution Field survey \citep[MOSDEF;][]{2015ApJS..218...15K} is one of the largest rest-frame optical spectroscopic samples of galaxies across the epoch of peak galaxy formation ($1 \lesssim z \lesssim 3$).
The exhaustive spectroscopic and multi-wavelength photometric data available for the MOSDEF sample enables the construction of a clean set of SFGs with no significant AGN contamination.
Furthermore, the sensitivity of the MOSDEF spectroscopy means that H$\alpha$ can be used as the reference SFR to probe the SFR-RC for more typical galaxies at $z\sim2$ than current FIR observations allow.
By combining the Balmer-decrement corrected H$\alpha$ SFR estimates of MOSDEF with the deepest available radio continuum observations we aim to directly calibrate the SFR-RC relation between $1.4 \lesssim z \lesssim 2.6$ without the use of confused far-infrared observations and any accompanying biases.
Specifically, we aim to check the validity of empirical SFR-RC relations calibrated in the local Universe and test for evidence of any redshift evolution.

The remainder of this paper is organized as follows:
In Section~\ref{sec:data}, we present details of the data used in this paper as well as details of the sample construction and AGN identification.
Section~\ref{sec:method} then describes the method and results for our analysis, alongside an investigation of the potential systematic uncertainties or biases in our measurements.
In Section~\ref{sec:discussion}, we present discussion of our results and the future prospects for radio luminosity derived SFRs in the next generation of radio continuum surveys. 
Finally, in Section~\ref{sec:summary} we summarise the results and our conclusions.
Throughout this paper, all magnitudes are quoted in the AB system \citep{1983ApJ...266..713O} unless otherwise stated. We also assume a $\Lambda$-CDM cosmology with $H_{0} = 70$ kms$^{-1}$Mpc$^{-1}$, $\Omega_{m}=0.3$ and $\Omega_{\Lambda}=0.7$.

\section{Data and Methodology}\label{sec:data}
\subsection{MOSDEF Survey}\label{sec:mosdef-data}
The sample we study in this work is drawn from the MOSFIRE Deep Evolution Field survey \citep[MOSDEF;][]{2015ApJS..218...15K}, an extensive rest-frame optical spectroscopic survey of $H_{160}$ magnitude-selected galaxies at $1.37 \leq z \leq 3.8$. 
In addition to providing the first representative samples of rest-frame optical spectra for galaxies on the star-forming main sequence in this epoch, the galaxies observed by MOSDEF have been drawn from the best studied deep extragalactic fields - enabling a wide variety of studies that exploit the extensive multi-wavelength ancillary data available.

We refer to \citet{2015ApJS..218...15K} for full details on the MOSDEF sample selection, observing strategy and data reduction procedures.
In summary, the primary magnitude cut for the two redshift windows used in this work, $1.37 \leq z \leq 1.7$ and $2.09 \leq z \leq 2.61$, was $H_{160}=24$ and 24.5 respectively.
These magnitude limits correspond to effective stellar mass limits of $\approx 10^{9}\textup{M}_{\odot}$ in each bin.
In this work we concentrate our analysis on the MOSDEF sources drawn from the CANDELS GOODS North field \citep{2011ApJS..197...35G,Koekemoer:2011br} due to the sensitivity of the radio continuum imaging available within the field.

H$\alpha$ and H$\beta$ line luminosities are measured through Gaussian fits to the observed line profiles of flux-calibrated spectra, with uncertainties on these luminosities estimated through Monte Carlo simulations \citep{2015ApJ...806..259R}. 
The observed line luminosities are then corrected for spectroscopic slit-losses using near-infrared photometry and the spectrum of a ``slit star'' placed on each observing mask \citep{2015ApJS..218...15K,2015ApJ...806..259R}, with uncertainties on these corrections of 16\% and 20\% for H$\alpha$ and H$\beta$, respectively. 
Additional corrections for underlying Balmer absorption are determined from the full UV to near-IR spectral energy distribution modelling of \citet{2015ApJ...806..259R}. 
Finally, dust corrections are applied to the line luminosities using the measured Balmer decrements (ratio of H$\alpha$ to H$\beta$) assuming the \citet{1989ApJ...345..245C} Galactic extinction curve \citep{2015ApJ...806..259R,Reddy2020,2015ApJ...815...98S}. 
H$\alpha$ SFRs are calculated from the dust-corrected H$\alpha$ luminosity following: 
\begin{equation}\label{eq:hasfr}
    \textup{SFR} (\text{M}_{\odot}\,\text{yr}^{-1})  = 10^{-41.257}\times L_{\text{H}\alpha} (\text{erg\,s}^{-1}),
\end{equation}
as presented in \citet[][]{2011ApJ...741..124H}, assuming a \citet{Kroupa:2001ki} initial mass function (IMF).
Throughout the paper, Eq.~\ref{eq:hasfr} is used to convert dust-corrected H$\alpha$ luminosity to SFR for the MOSDEF sample, as well as for literature relations for the SFR-RC relation (Table~\ref{tab:sfr-rc}). 

\subsection{Radio continuum imaging}\label{sec:radio-data}
Radio continuum flux measurements for the GOODS-N MOSDEF sample are extracted from sensitive \emph{Karl G. Jansky} Very Large Array (VLA) observations at 1-2 GHz of \citet[][with a central frequency of 1.525GHz]{2018ApJS..235...34O}.
These observations reach an rms noise of 2.2\ujy at the field centre, with a full-width half maximum (FWHM) resolution of 1.6\arcsec.
We first cross-match the catalog of blind radio source detections with the full 3D-HST source catalog in the field \citep[][from which the MOSDEF source positions are defined]{Skelton:do} with a large matching radius of 2$\arcsec$.
From the distribution of offsets between the two samples we measure a small average astrometric offset in $\Delta\alpha =  0.22\arcsec$ in R.A. and $\Delta\delta = 0.05\arcsec$ in Declination (c.f. the radio image pixel size of 0.35$\arcsec$).
After correcting for the small positional offsets between the two datasets, robust source identifications for the blind radio catalog are then selected as sources with a 3D-HST counterpart within a reduced matching radius of 0.8$\arcsec$.

Radio continuum flux density measurements for sources not detected in the blind catalogs are derived from measurements of the peak flux density, $S_{\text{Peak}}$, at the optical position of individual MOSDEF sources in the primary beam corrected radio continuum image (with optical positions corrected to the reference frame of the radio data as above).
Under the assumption that all sources are unresolved at the resolution of the radio imaging, the integrated flux density is defined as $S_{\text{Int}} = S_{\text{Peak}}$.
Corresponding flux density uncertainties are calculated based on the local noise around each source (within a $30\arcsec\times30\arcsec$ region).
Comparing our MOSDEF prior-driven (or `forced') peak measurements with the integrated flux measurements for unresolved radio sources in the blind catalog, we find a median $S_{\text{Int}}/S_{\text{Peak}} = 1.07$, with $\approx90\%$ of measurements within 1$\sigma$ agreement.
Given this close agreement, we choose not to apply any peak-to-total flux corrections to our forced radio flux measurements (the potential effects on our results of this assumption are discussed in Section~\ref{sec:systematics}).
Finally, to allow direct comparison between our radio measurements and literature measurements of the SFR-radio relation at $z\sim0$, we convert our measured fluxes and luminosities from their central frequency of 1.525\,GHz to 1.4\,GHz assuming a spectral slope of $\alpha=-0.8$ \citep[a valid assumption for RC emission produced by star-formation,][]{2017MNRAS.469.3468C,2019arXiv190307632G}.

\subsection{Sample selection}\label{sec:sample}
We restrict the full MOSDEF spectroscopic sample to sources in GOODS-N in the redshift range $1.37 < z < 2.61$, requiring measurements of both H$\alpha$ and H$\beta$.
With the additional requirement that H$\alpha$ luminosity is measured to $>3\sigma$, our sample consists of a total of 178 sources.
Of this sample, 63 sources are not detected in H$\beta$ at the $3\sigma$ required for constraints on the dust-corrected H$\alpha$ SFR.
Radio continuum detections of $>2\sigma$ are found for 55/178 of the full MOSDEF sample and 35/114 sources with H$\alpha$ SFR estimates.

\begin{figure}
\centering
 \includegraphics[width=0.95\columnwidth]{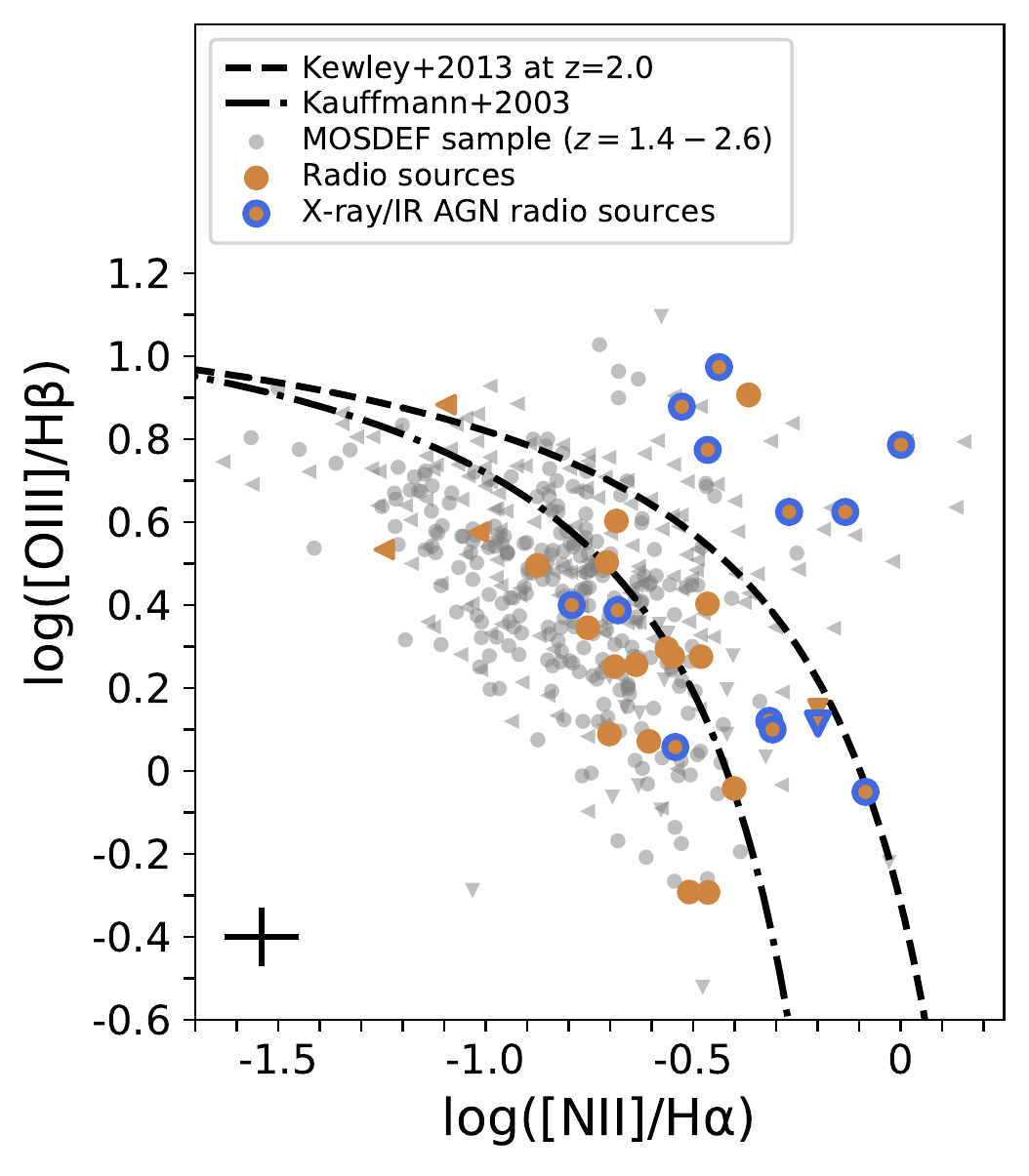}
  \caption{Rest-frame optical emission line diagnostics \citep[or BPT diagram;][]{1981PASP...93....5B} for the radio-detected MOSDEF sources (orange points) and the main MOSDEF sample (gray points). Sources with formal detections in all lines are plotted as circles, while non-detections in either [O\textsc{iii}] or [N\textsc{ii}] are plotted as triangles indicating $3\sigma$ upper limits. Robust X-ray or IR selected AGN are illustrated by blue edged circles and triangles.
  Also shown are the optical AGN classification criteria of \citet[][dashed-dotted line]{2003MNRAS.346.1055K} 
  and the redshift dependent criteria of \citet[][dashed line; illustrated at $z=2$]{2013ApJ...774L..10K}.}
 \label{fig:bpt}
\end{figure}





To produce as clean a sample of star-forming galaxies as possible, we combine all the available multi-wavelength information to identify AGN activity, including X-ray emission, mid-IR colours and optical line diagnostics and kinematic information \citep{2015ApJ...801...35C,2017ApJ...835...27A,2018ApJ...866...63A, 2019ApJ...886...11L}.
Of the radio detected sources, we find that 19 sources are identified as an X-ray AGN based on the \emph{Chandra} Deep Field North observations of \citet[][see also \citeauthor{2015MNRAS.451.1892A}~\citeyear{2015MNRAS.451.1892A}]{2003AJ....126..539A}.
Of these X-ray AGN sources, four are also identified as an AGN based on their mid-infrared colours satisfying the criteria of \citet{Donley:2012ji}.
An additional two sources satisfy the IR AGN selection criteria but are not X-ray detected.

For the purposes of optical AGN classification, the optical lines are fitted with a narrow and a broad Gaussian component, representing the narrow- and broad-line emission from the AGN and/or the host galaxy, respectively, and an additional Gaussian profile with a velocity offset to represent a potential blueshifted outflow component, as described in Section 2.4 of \citet{2017ApJ...849...48L}.
In Fig.~\ref{fig:bpt} we show the resulting \citet[][BPT]{1981PASP...93....5B} diagram for the full sample of MOSDEF sources with measurements in all four of the H$\beta$, [O{\sc iii}]$\lambda 5008$, H$\alpha$, and [N{\sc ii}]$\lambda 6585$ emission lines in the redshift range of interest (gray points),
as well as the radio-selected sample within GOODS North (brown points).
Overall, we see that the radio detected sample covers a very similar parameter space to the full MOSDEF sample but that a large proportion of the radio sources with BPT measurements are also identified as X-ray/IR AGN. 

Based on the BPT emission line diagnostics we identify an additional three robust optical AGN based on $\log([\textsc{Nii}]/\text{H}\alpha) > -0.3$ that are not identified by X-ray/IR AGN selection criteria (note that sources may not be represented in Fig.~\ref{fig:bpt} if H$\beta$ or [O{\sc iii}] measurements are not present).
The combined sample of 24 X-ray, mid-IR and $\log([\textsc{Nii}]/\text{H}\alpha)$ selected sources comprise our robust AGN sample.
A further seven MOSDEF sources are found to have optical line ratios that fall between the $z\sim0$ AGN/SF boundary defined by \citet[][dashed-dotted line]{2003MNRAS.346.1055K} and the maximum line ratios for a star-forming galaxy \citep[][dashed line]{2013ApJ...774L..10K} at the corresponding redshift, indicative of possible so-called `composite' sources \citep{2013ApJ...774..100K}.

The exact cause of the elevated $\log([\textsc{Oiii}]/\text{H}\beta)$ line ratios in these sources is a subject of debate within the literature, with some studies identifying AGN activity as the cause \citep[e.g.][]{2014ApJ...781...21N} and others finding that evolution in the stellar ionizing radiation, ionization parameter and metal abundances within star-forming galaxies can account for the observed line-ratios \citep{2014ApJ...785..153M,2014ApJ...795..165S,2016ApJ...826..159S,2016ApJ...816...23S,2019ApJ...881L..35S,2017ApJ...836..164S,2020MNRAS.495.4430T}.
Within the MOSDEF sample specifically, \citet{2015ApJ...801...35C} illustrate that many of the MOSDEF sources that lie in this `composite' region are likely star-forming galaxies and not AGN.
We therefore keep these sources within our sample during our analysis, however in Section~\ref{sec:sfr_stack} we discuss what effect these sources may have on our results.
All 24 of the robustly identified AGN are excluded from our subsequent analysis of individual radio detections (and all 34 robust AGN within the full GOODS-N sample are excluded from any stacking analysis outlined below).

\subsection{Stacking of radio continuum imaging}\label{sec:rc_stack}
Stacking analysis represents a powerful tool for setting meaningful constraints on population properties far below the nominal detection threshold for a given dataset.
Parametric stacking methods \citep[e.g.][]{Roseboom:2014jw,Zwart:2015kk} have the potential for constraining not just average properties within a sample, but precise measurements of the source count distributions or luminosity functions.
However, given the small samples currently available to us and the simple question being asked, we take a more simple median stacking approach. 

For a given sample, the stacked radio flux measurement is obtained by taking the median of the distribution of peak flux density at the position of every source within the radio map.
As in Section~\ref{sec:radio-data}, this assumes that the stacked population are unresolved at the resolution of the JVLA radio imaging \citep[$\text{FWHM} = 1.6\arcsec$, an assumption supported by high-resolution imaging of radio sources within the field;][]{2020MNRAS.495.1188M}.
We estimate uncertainties on the stacked flux measurements by calculating the standard deviation of the median flux from jackknife re-sampling of the peak pixel values before stacking.
Additionally, we estimate the stacked background noise by taking the robust standard deviation in a 2D stack of $30\arcsec\times30\arcsec$ region around each source.
For our total flux uncertainty we combine the two measurements in quadrature, i.e. $\sigma_{\text{Tot}}^{2} = \sigma_{\text{Sample}}^{2} + \sigma_{\text{RMS}}^{2}$.
The derived flux measurement and corresponding uncertainty are converted to a luminosity using the median spectroscopic redshift of the sample.

\subsection{Stacking of rest-frame optical spectra}\label{sec:opt_stack}
To include sources that are individually undetected in H$\beta$ in our optical analysis, we stack the near-IR spectra as follows\footnote{Source code available at \url{https://github.com/IreneShivaei/specline}, \citet{2018ApJ...855...42S}}. 
Individual spectra are shifted to the rest frame with a uniform wavelength grid of $\Delta \lambda=0.5$\,{\AA}. 
The composite spectrum is taken as the median luminosity at a given wavelength bin while the composite error spectrum is defined as ${1}/{\sqrt{\sum\limits_{i} \sigma_i^2}}$, where $\sigma_i^2$ is the variance of the $i$th individual spectrum. 
We choose to use the median values for consistency with the stacked radio measurements.
However, we note that the median and mean composite spectra are found to provide fully consistent results and the stacked line measurements are consistent with previous MOSDEF studies that make use of mean luminosities \citep{2015ApJ...815...98S,2018ApJ...855...42S}.

The stacked H$\beta$ and H$\alpha$ lines are fit with single Gaussian profiles. 
The composite line luminosities and errors are estimated as the mean and standard deviation of 1000 Monte Carlo simulations, where the composite spectrum is randomly perturbed within the error spectrum. 
The H$\alpha$ and H$\beta$ emission lines are further corrected for underlying Balmer absorption using the median absorption of individual galaxies estimated from SED fitting, inversely weighted by individual H$\alpha$ and H$\beta$ flux uncertainties, respectively.

To derive a composite Balmer decrement ($\langle$H$\alpha$/H$\beta\rangle$), individual spectra are initially normalised to their H$\alpha$ luminosities and the normalised composite H$\beta$ line is fit by a Gaussian profile. The Balmer decrement is the inverse value of the composite $\langle$H$\beta$/H$\alpha\rangle$.
For additional details regarding the stacking technique for rest-optical spectra we refer the reader to Section 3.1 of \citet{2018ApJ...855...42S}.

\section{The SFR-RC relation at $1.4 < \lowercase{z} < 2.6$}\label{sec:method}
\subsection{RC Luminosity as a function of H$\alpha$ SFR for individual galaxies}
\begin{figure}
\centering
 \includegraphics[width=\columnwidth]{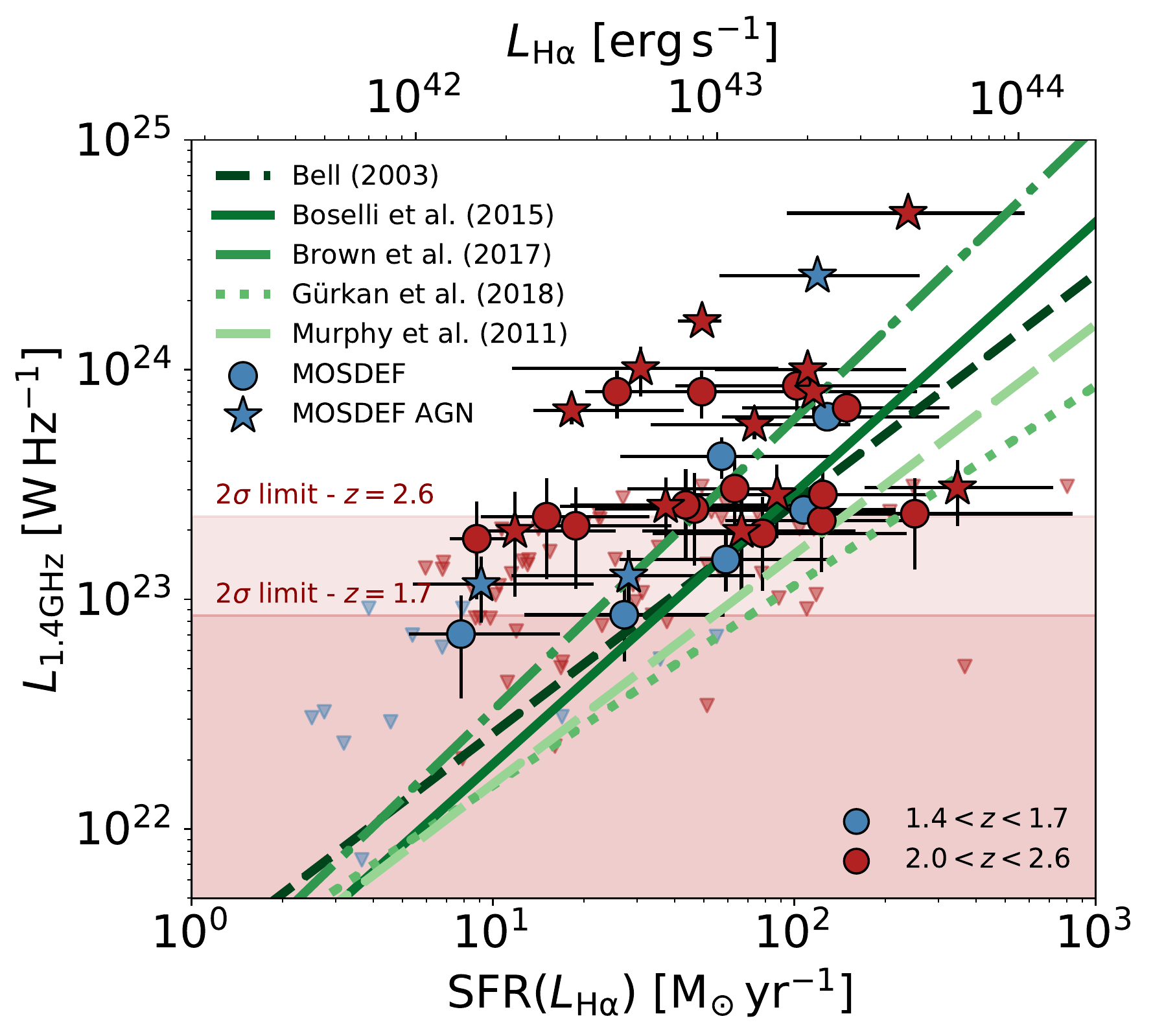}
  \caption{Measured distribution of SFRs vs radio continuum luminosities (filled circles/stars), or $1\sigma$ upper limits (triangles), for our full MOSDEF sample. The colour scale for the observed data corresponds to the measured spectroscopic redshifts. Empirical $z\sim0$ relations for the SFR-RC relation are plotted for reference as the green lines - see Table~\ref{tab:sfr-rc}. Sources identified as AGN through optical, IR and X-ray criteria are also shown for reference (coloured stars). $L_{\text{H}\alpha}$ measurements are converted to SFR following Equation.~\ref{eq:hasfr}. 
  }
 \label{fig:sfr_rc}
\end{figure}

In Fig.~\ref{fig:sfr_rc} we plot the observed distribution of dust-corrected SFR versus 1.4 GHz radio luminosity ($L_{\textup{1.4GHz}}$, in units of W Hz$^{-1}$) for our radio and H$\alpha$+H$\beta$ detected sample.
For H$\alpha$+H$\beta$ sources undetected in radio, we also plot $1\sigma$ upper limits.
Alongside the observed high-redshift measurements, we show five published empirical SFR-RC relations from the literature (summarised in Table~\ref{tab:sfr-rc}).
While we find that our high redshift observations are broadly consistent with the $z\sim0$ relations, the main locus of datapoints appears to lie above the $L_{\textup{1.4GHz}}$ predicted from star-formation.
However, despite being the deepest data currently available, the radio continuum observations are only sensitive enough to robustly detect the higher SFR objects within the MOSDEF sample.
The observed SFR-radio relation therefore potentially suffers substantially from Malmquist-like bias, whereby the radio detected sample includes only sources that are scattered above the SFR-RC relation (and hence above the detection threshold).

\begin{table}
	\centering
	\caption{Literature empirical observational relations between the dust-corrected H$\alpha$ luminosity, $L_{\text{H}\alpha}$, and radio continuum luminosity, $L_{\textup{1.4GHz}}$, at $z\sim0$. SFRs have been converted to $L_{\text{H}\alpha}$ based on the appropriate initial mass-function for a given paper  \citep[see Table~3 of][]{Brown:2017uu}.}
	\label{tab:sfr-rc}
	\setlength{\tabcolsep}{1pt} 
	\renewcommand{\arraystretch}{2.0}
	\begin{tabular}{ll} 
		\hline
		Reference & $z\sim0$ Relation \\
		\hline
		\citet{Bell:2003bj} & $\log_{10} \left ( \frac{L_{\textup{1.4GHz}}}{\text{W\,Hz}^{-1}} \right ) =  \log_{10}\left(\frac{L_{\text{H}\alpha}}{10^{40} \text{erg\,s}^{-1}} \right ) + 20.16$ \\
		{\citet{2015A&A...579A.102B}} & $\log_{10} \left ( \frac{L_{\textup{1.4GHz}}}{\text{W\,Hz}^{-1}} \right ) =  1.18 \log_{10}\left(\frac{L_{\text{H}\alpha}}{10^{40} \text{erg\,s}^{-1}} \right ) + 19.62$ \\
		\citet{Brown:2017uu} & $\log_{10} \left ( \frac{L_{\textup{1.4GHz}}}{\text{W\,Hz}^{-1}} \right ) =  1.27\log_{10}\left(\frac{L_{\text{H}\alpha}}{10^{40} \text{erg\,s}^{-1}} \right ) + 19.65$ \\
		\citet{2018MNRAS.475.3010G} & $\log_{10} \left ( \frac{L_{\textup{1.4GHz}}}{\text{W\,Hz}^{-1}} \right ) = 0.87 \log_{10}\left(\frac{L_{\text{H}\alpha}}{10^{40} \text{erg\,s}^{-1}} \right ) + 20.51$ \\
		\citet{2011ApJ...737...67M} & $\log_{10} \left ( \frac{L_{\textup{1.4GHz}}}{\text{W\,Hz}^{-1}} \right ) =  \log_{10}\left(\frac{L_{\text{H}\alpha}}{10^{40} \text{erg\,s}^{-1}} \right ) + 19.94$ \\
		\hline
	\end{tabular}
\end{table}

When comparing our results to a range of literature parametrisations of the SFR-RC relation at $z\sim0$ \citep[or $z \lesssim 0.3$ in the case of][]{2018MNRAS.475.3010G}, we find a large range of predicted $L_{\text{1.4\,GHz}}$ at the highest SFR.
At $\text{SFR} \sim 200\, \text{M}_{\odot}\,\text{yr}^{-1}$ the predicted $L_{\text{1.4\,GHz}}$ span approximately an order of magnitude - far in excess of the typical scatter measured within individual studies \citep[$\approx0.2$ to 0.3 dex][]{Bell:2003bj,Brown:2017uu}.
The reason for this diversity is likely due to the combined biases of both observational limitations (i.e. small volumes limiting the number of rarer high SFR galaxies) and evolutionary effects (i.e. the decline in the overall cosmic SFR density) that restrict the $z\sim0$ analysis to much smaller SFR than typically probed at $z > 1$.
For example, in those studies where the slope of the SFR-RC has been fit as a free parameter \citep{Brown:2017uu,2018MNRAS.475.3010G}, the samples are generally limited to $\text{SFR} \lesssim 30\, \text{M}_{\odot}\,\text{yr}^{-1}$.

Below $\text{SFR} \sim 30\, \text{M}_{\odot}\,\text{yr}^{-1}$, the $z\sim0$ relations presented in Table~\ref{tab:sfr-rc} all predict $L_{\text{1.4\,GHz}}$ that are within the 0.2 dex intrinsic scatter.
Given the observational limits of the RC data demonstrated in Fig.~\ref{fig:sfr_rc}, sub-threshold analysis or stacking measurements are required to make meaningful constraints on the SFR-RC relation in the regime where $z\sim0$ SFR-RC relations are well calibrated ($\text{SFR} \lesssim 100\, \text{M}_{\odot}\,\text{yr}^{-1}$).

\subsection{Stacked RC Luminosity as a function of H$\alpha$ SFR}\label{sec:sfr_stack}
We first explore the stacked radio luminosities as a function of dust-corrected H$\alpha$ derived SFR.
Following the method described in Section~\ref{sec:rc_stack}, we stack the radio imaging for bins of SFR to measure the corresponding average 1.4 GHz radio luminosity.
At $2.0 < z < 2.6$ we are able split our sample into five bins of SFR spanning over two orders of magnitude. 
However, at $1.4 < z < 1.7$, the smaller available sample size limits the analysis to only two bins in SFR.

\begin{figure}
\centering
 \includegraphics[width=\columnwidth]{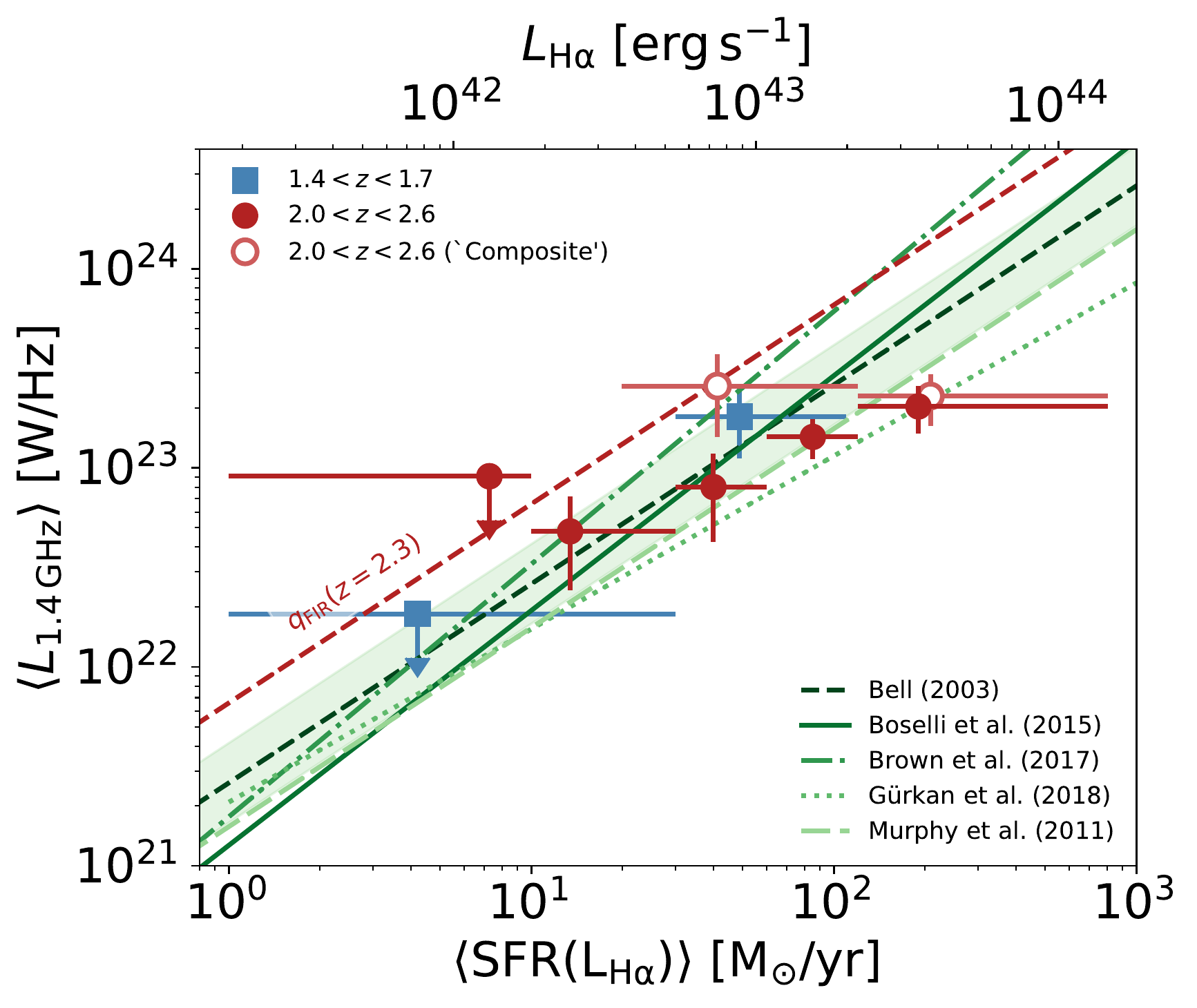}
  \caption{Median stacked radio luminosities for bins of dust-corrected H$\alpha$ SFR at $1.4< z < 1.7$ (blue squares) and $2.0 < z < 2.6$ (red filled circles). Stacks for the subset classified as `composite' in the BPT diagram are plotted as red open circles for comparison. Empirical $z\sim0$ relations for the SFR-RC relation are plotted for reference as the green lines - see Table~\ref{tab:sfr-rc}. We show the \citet{Bell:2003bj} SFR-radio relation scaled to $z=2.3$ based on the observed evolution of the far-infrared radio correlation from \citet[][red dashed line]{2017MNRAS.469.3468C}. $L_{\text{H}\alpha}$ measurements are converted to SFR following Equation.~\ref{eq:hasfr}.
  }
 \label{fig:stacks_sfrbin}
\end{figure}

Fig.~\ref{fig:stacks_sfrbin} presents the median radio luminosity as a function of dust-corrected H$\alpha$ for the H$\alpha$ and H$\beta$ detected MOSDEF sample (with the corresponding values also provided in Table~\ref{tab:sfr_stack}).
As in Fig.~\ref{fig:sfr_rc} we present a range of literature relations for the SFR-RC for comparison.
In addition, we also illustrate the predicted redshift evolution of the \citet{Bell:2003bj} relation based on the observed evolution of the far-infrared radio correlation \citep[as measured by][]{2017MNRAS.469.3468C}.

We find that our observed radio luminosities at given SFR are fully consistent with the expectations from $z\sim0$ relations. 
At $2.0 < z < 2.6$, where our S/N is highest, we find that our measurements favour a flatter SFR-RC relation with our results statistically inconsistent with the steeper relation measured by \citet{Brown:2017uu}.
However, given the intrinsic scatter typically found for measurements of the SFR-RC relation ($\sim0.2$ dex) and the small sample sizes available, we cannot place meaningful constraints on the high-redshift SFR-RC slope.

As referenced in Section~\ref{sec:sample}, the exact nature of $z > 1$ sources classified as `composite' sources in the BPT diagnostic is somewhat unknown.
To see what effect these sources may be having on our stacked SFR-RC relation, we repeat the stacking analysis using only the subset of `composite' MOSDEF sources in the $2.0 < z < 2.6$ bin (open circles in Fig.~\ref{fig:stacks_sfrbin}).

At high SFR ($>120\,\text{M}_{\odot}\,\text{yr}^{-1}$, 4 sources) the composite sample is in excellent agreement with the radio luminosity measured for the full sample at similar SFR.
However, we find that the radio luminosity for the lower SFR composite bin ($<120\,\text{M}_{\odot}\,\text{yr}^{-1}$, 9 sources) is significantly higher than that measured for full sample at similar SFR - potentially indicative of low-level AGN activity within the sample.
The excess in radio emission for this datapoint places it in better agreement with the SFR-RC relation predicted from the FIRRC \citep[$L_{\textup{IR}}/L_{\textup{1.4GHz}}\propto (1+z)^{-0.15}$;][]{2017MNRAS.469.3468C}.
However, given the large uncertainties, it is still formally consistent within $2\sigma$ of many of the $z\sim0$ SFR-RC relations and the measurements for the full sample.
Additionally, the potential AGN contribution to the H$\alpha$ and H$\beta$ emission lines means that our inferred H$\alpha$ SFR may be biased for some individual sources.

We note that the RC luminosities are consistent with the conclusions of \citet{2015ApJ...801...35C}, whereby the sources lying between the \citet{2003MNRAS.346.1055K} and \citet{2001ApJ...556..121K} lines on the BPT diagram may have both star-formation and AGN contributions to the emission lines, under the assumption that the excess radio continuum emission is a result of AGN activity.
However, given the numerous correlations between interstellar medium (ISM) properties and the RC emission produced for a given SFR \citep{2010ApJ...717....1L}, the evolving ISM conditions that give rise to elevated $[\textsc{Oiii}]/\text{H}\beta$ and $\log([\textsc{Nii}]/\text{H}\alpha)$ line ratios could also result in the observed RC excess \citep{2016ApJ...826..159S, 2019ApJ...881L..35S}.
Due to the small samples available for this study and the low statistical significance of the observed excess, we cannot make any strong conclusions regarding the properties of the composite BPT population at these redshifts based on the observed RC properties.
Nevertheless, these results demonstrate that our calibration of the SFR-RC relation at these redshifts is robust to potential contamination from a small number of residual AGN within the sample.

\begin{table*}
	\centering
	\caption{Measured properties of the primary SFR selected samples in Fig.~\ref{fig:stacks_sfrbin}. For each redshift and SFR range, we present the sample size, average SFR within the bin and the corresponding radio luminosity. $L_{\text{H}\alpha}$ measurements are converted to SFR following Equation.~\ref{eq:hasfr}. Limits on $L_{\text{1.4\,GHz}}$ represent 2$\sigma$ upper limits based on the measured background noise in the 2D median stack.}
	\label{tab:sfr_stack}
\begin{tabular}{ccccc}
\hline
$\textup{SFR}(L_{\text{H}\alpha})$ Range -$\phi$ ($\text{M}_{\odot}\,\text{yr}^{-1}$) & $N$ & $\left \langle 
\textup{SFR}(L_{\text{H}\alpha})\right \rangle$ ($\text{M}_{\odot}\,\text{yr}^{-1}$) &  $\left \langle L_{\text{H}\alpha} \right \rangle $ ($10^{42} \text{erg\,s}^{-1}$) & $\left \langle L_{\text{1.4\,GHz}} \right \rangle$ ($10^{22} \text{W\,Hz}^{-1}$) \\
\hline
\multicolumn{5}{c}{$1.4 < z < 1.7$}\\
\hline
$1 < \phi \leq 30$ & 14 & 4.2 & $0.8$ & $<1.8$ \\
$30 < \phi$ & 6 & 48.8 & $8.8$ & $18.1 \pm 7.0$ \\
\hline
\multicolumn{5}{c}{$2.0 < z < 2.6$}\\
\hline
$1 < \phi \leq 10$ & 10 & 7.3 & $1.3$ & $<9.1$ \\
$10 < \phi \leq 30$ & 25 & 13.4 & $2.4$ & $4.8 \pm 2.4$ \\
$30 < \phi \leq 60$ & 17 & 40.0 & $7.2$ & $8.0 \pm 3.8$ \\
$60 < \phi \leq 120$ & 13 & 85.2 & $15.4$ & $14.3 \pm 3.3$ \\
$120 < \phi$ & 8 & 190.4 & $34.4$ & $20.4 \pm 5.5$ \\
\hline
\end{tabular}
\end{table*}

\subsection{Stacked RC and H$\alpha$ SFR as a function of stellar mass}\label{sec:mass_stack}
To measure the Balmer decrement for individual sources in the MOSDEF spectra, we require a $3\sigma$ detection in both H$\alpha$ and H$\beta$ emission lines.
However, the requirement of a H$\beta$ detection potentially biases the SFR selected samples to sources with lower dust corrections and further limits the available samples for stacking measurements within the radio data.

Stellar mass represents an ideal parameter for stacking in a number of ways.
Firstly, given the sensitivity and breadth of the available photometry in the GOODS-N field, robust measurements of stellar mass are available for the full MOSDEF sample regardless of SFR or emission line S/N.
Secondly, the stellar mass estimates themselves are derived from photometric data independent of the two datasets being stacked.
Finally, thanks to the strong correlation between stellar mass and star-formation rate across all redshifts \citep[e.g.][]{Noeske:2007kq,2011ApJ...738..106W,Whitaker:2014fs,2015ApJ...815...98S}, stellar mass itself represents a good proxy for star-formation and allows us to maintain a wide dynamic range within our stacked measurements.

We make use of stellar mass estimates for the MOSDEF sample from best-fit SED models as described in \citet[][see Section 4.5]{2015ApJS..218...15K}.
For the MOSDEF sample used in our analysis, the quoted statistical uncertainty on individual stellar mass estimates is typically 0.18 dex.
Note that additional systematic uncertainties in inferring stellar masses from broadband photometry mean that the quoted uncertainties likely underestimate the true stellar mass uncertainty \citep[see e.g.][]{Conroy:2009ks,Santini:2015hh}\footnote{We also note that the stellar mass estimates assume a \citet{2003PASP..115..763C}
IMF, which is similar but not identical to the \citet{Kroupa:2001ki} IMF assumed throughout the rest of this work. However the difference in mass to light ratios is not signficant compared to the uncertainty on individual masses.}.
However, as the stellar mass values simply represent an additional parameter by which to appropriately bin our sample for stacking, absolute precision in the estimates are not important for our analysis.
Nevertheless, the uncertainties are small relative to the bin size and the average stellar mass probed per stack is not significantly affected by the error on individual measurements.

\begin{figure}
\centering
 \includegraphics[width=\columnwidth]{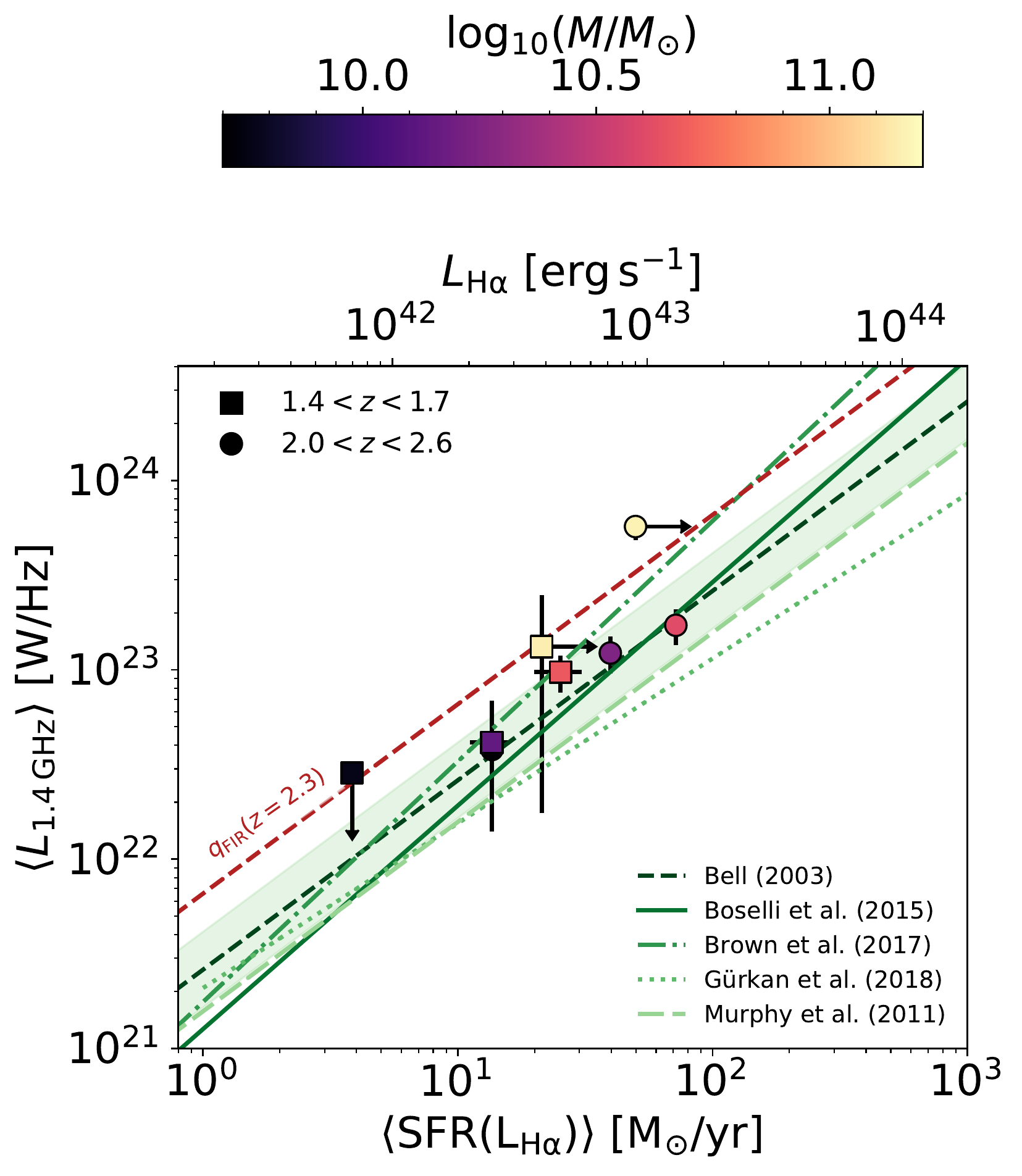}
  \caption{Median stacked radio luminosity versus median-stacked dust-corrected H$\alpha$ SFR for bins of galaxy stellar mass  at $1.4< z < 1.7$ (squares) and $2.0 < z < 2.6$ (circles). Datapoints are coloured based on the median stellar mass within the sample (see text for details). Literature SFR-radio relations at $z\sim0$ and the predicted $z\sim2.3$ relation are plotted as in Fig.\ref{fig:stacks_sfrbin}. $L_{\text{H}\alpha}$ measurements are converted to SFR following Equation.~\ref{eq:hasfr}.}
 \label{fig:stacks_massbin}
\end{figure}

In Fig.\ref{fig:stacks_massbin} we present the radio luminosity as a function of Balmer-decrement corrected $L_{\rm {H\alpha}}$ for four bins of stellar mass in each redshift range.
The corresponding data are also presented in Table~\ref{tab:mass_stack}. 
For the highest mass bin at $2.0 < z < 2.6$, the H$\beta$ emission line is not detected with any statistical significance.
We therefore derive a \emph{lower} limit on the Balmer decrement by correcting for dust extinction using the 2$\sigma$ upper limit on H$\beta$.

We find our results based on mass-selected samples to be fully consistent with the stacking measurements based on individual SFR measurements (Fig.~\ref{fig:stacks_sfrbin}) - with the measured radio luminosities fully within the 1$\sigma$ of the intrinsic scatter for $z\sim0$ SFR-RC relations.
Additionally, we find that our observations at $1.4< z < 1.7$ (squares) and $2.0 < z < 2.6$ (circles) are in excellent agreement with no evidence for redshift evolution within the sample.
As with Fig.~\ref{fig:stacks_sfrbin}, we find that our observations are in better agreement with $z\sim0$ relations that exhibit flatter slopes.

For the range of stellar masses explored in our sample, we see no evidence for any effect of stellar mass on the SFR-RC relation.
The SFR measured as a function of mass for each redshift bin are also consistent within the uncertainties with literature parametrisations of the stellar mass-SFR relation over these redshifts \citep[see e.g.][]{2014ApJS..214...15S} and with measurements from the parent MOSDEF sample \citep{2015ApJ...815...98S}.

\begin{table*}
	\centering
	\caption{Measured properties of the stellar mass selected samples in Fig.~\ref{fig:stacks_massbin}. For each redshift and mass range, we present the average stellar mass and sample size along with the stacked measurements. H$\alpha$ line luminosities are Balmer decrement corrected following the method described in the text. $L_{\text{H}\alpha}$ measurements are converted to SFR following Equation.~\ref{eq:hasfr}. Limits on $L_{\text{H}\alpha}$ or $L_{\text{1.4\,GHz}}$ represent 2$\sigma$ lower and upper limits respectively.}
	\label{tab:mass_stack}
\begin{tabular}{cccccc}
\hline
Stellar Mass Range & Median Mass ($\log_{10}(\text{M/M}_{\odot})$) & $N$ & $\left \langle L_{\text{H}\alpha} \right \rangle $ ($10^{42} \text{erg\,s}^{-1}$) & $\left \langle 
\textup{SFR}(L_{\text{H}\alpha})\right \rangle$ ($\text{M}_{\odot}\,\text{yr}^{-1}$) & $\left \langle L_{\text{1.4\,GHz}} \right \rangle$ ($10^{22}\, \text{W\,Hz}^{-1}$) \\
\hline
\multicolumn{6}{c}{$1.4 < z < 1.7$}\\
\hline
$9 < M/M_{\odot} < 10$ & 9.77 & 17 & $0.7 \pm 0.1$ & $3.8 \pm 0.3$ & $<2.9$ \\
$10 < M/M_{\odot} < 10.5$ & 10.14 & 7 & $2.5 \pm 0.4$ & $13.6 \pm 2.4$ & $4.1 \pm 2.7$ \\
$10.5 < M/M_{\odot} < 11.0$ & 10.67 & 4 & $4.6 \pm 1.0$ & $25.3 \pm 5.4$ & $9.7 \pm 2.1$ \\
$11 < M/M_{\odot} < 11.5$ & 11.15 & 2 & $>3.9$ & $>21.3$ & $13.3 \pm 11.5$ \\
\hline
\multicolumn{6}{c}{$2.0 < z < 2.6$}\\
\hline
$9 < M/M_{\odot} < 10$ & 9.7 & 57 & $2.5 \pm 0.1$ & $13.6 \pm 0.7$ & $3.8 \pm 2.2$ \\
$10 < M/M_{\odot} < 10.5$ & 10.25 & 35 & $7.2 \pm 0.3$ & $39.7 \pm 1.8$ & $12.3 \pm 2.7$ \\
$10.5 < M/M_{\odot} < 11.0$ & 10.61 & 16 & $13.0 \pm 0.6$ & $71.7 \pm 3.6$ & $17.2 \pm 3.7$ \\
$11 < M/M_{\odot} < 11.5$ & 11.16 & 4 & $>9.0$ & $>49.8$ & $57.1 \pm 8.6$ \\
\hline
\end{tabular}
\end{table*}

\subsection{Systematic effects on the SFR-RC}\label{sec:systematics}
The assumptions we made in this study could have a systematic effect on both the normalisation and the slope of the measured SFR-RC.
In this section we discuss these assumptions and their potential effects on the results presented in Sections~\ref{sec:sfr_stack} and \ref{sec:mass_stack}.

Firstly, in deriving dust corrections from the measurements of Balmer decrement we have assumed a \citet{1989ApJ...345..245C} Galactic extinction curve.
Multiple studies have shown good agreement between dust-corrected H$\alpha$ SFR assuming the \citeauthor{1989ApJ...345..245C} curve and SFR independently derived from IR data \citep[e.g.][]{2016ApJ...820L..23S}, while \citet{Reddy2020} explicitly show that the nebular attenuation curve relevant for MOSDEF galaxies is similar to that of \citeauthor{1989ApJ...345..245C}.
Nevertheless, a change in the assumed dust attenuation curve could systematically change the SFR-RC relation presented in this work.
Correcting for dust attenuation assuming a \citet{2000ApJ...533..682C} attenuation curve leads to an increase in stacked MOSDEF SFR estimate for the $10.5 < \log_{10}(\text{M/M}_{\odot}) < 11.0$ mass bin of 0.05 and 0.07 dex at $1.4 < z < 1.7$ and $2.0 < z < 2.6$ respectively (with smaller offsets at lower mass).
Our stacked dust-corrected H$\alpha$ SFR would need to be lower by $\sim0.4$ dex on average to bring the observed SFR-RC relation in line with predictions based on the evolving FIRRC.
We can therefore be confident that the key conclusion presented in this work is not affected by the choice of dust attenuation curve.

Secondly, we assume a single $L({\rm H\alpha})$-to-SFR conversion for the entire sample (Section~\ref{sec:method}), irrespective of the effect of stellar metallicity and binarity on the conversion factor \citep[e.g.][]{2016MNRAS.456..485S,2018ApJ...869...92R,2019ApJ...871..128T,2019MNRAS.490.5359W}. When using stellar population synthesis models that include binaries, the $L({\rm H\alpha})$-to-SFR conversion factor varies by $\sim 0.3$\,dex from stellar metallicity of $Z_*=0.2$ to 1.0\,$Z_{\odot}$. 
Based on observations of the mass-metallicity relation out to $z > 2$ \citep{2006ApJ...644..813E,2014MNRAS.440.2300C,2015ApJ...799..138S}, a sub-solar conversion factor might be more applicable to the lowest mass bin in Figure 4, shifting the H$\alpha$ SFRs to lower values by up to $\sim 0.3$\,dex. 
However, our observed relations between H$\alpha$ luminosities (i.e. upper axes in Fig.\ref{fig:sfr_rc}, \ref{fig:stacks_sfrbin} and \ref{fig:stacks_massbin}) and radio continuum luminosities stay unaffected by such systematic uncertainties.

Thirdly, in deriving the radio luminosity measurements we have assumed a constant canonical spectral slope of $\alpha = -0.8$ \citep{1992ARA&A..30..575C} when converting from flux to luminosity (or vice versa) throughout this work. 
Although this canonical value has been shown to be a valid assumption for the median spectral slope in SFGs out to high redshift \citep{2017MNRAS.469.3468C,2019arXiv190307632G}, there is still significant scatter around this value. 
\citet{2019arXiv190307632G} illustrate that not accounting for this intrinsic variation can result in both significantly higher scatter in the FIRRC and a potential bias.
However, our results are unlikely to be significantly affected by this scatter due to the use of stacked samples, while the magnitude bias measured by \citet[][$\Delta q_{\textup{FIR}}= -0.061$]{2019arXiv190307632G} is not large enough to affect our conclusions.

\begin{figure}
\centering
 \includegraphics[width=\columnwidth]{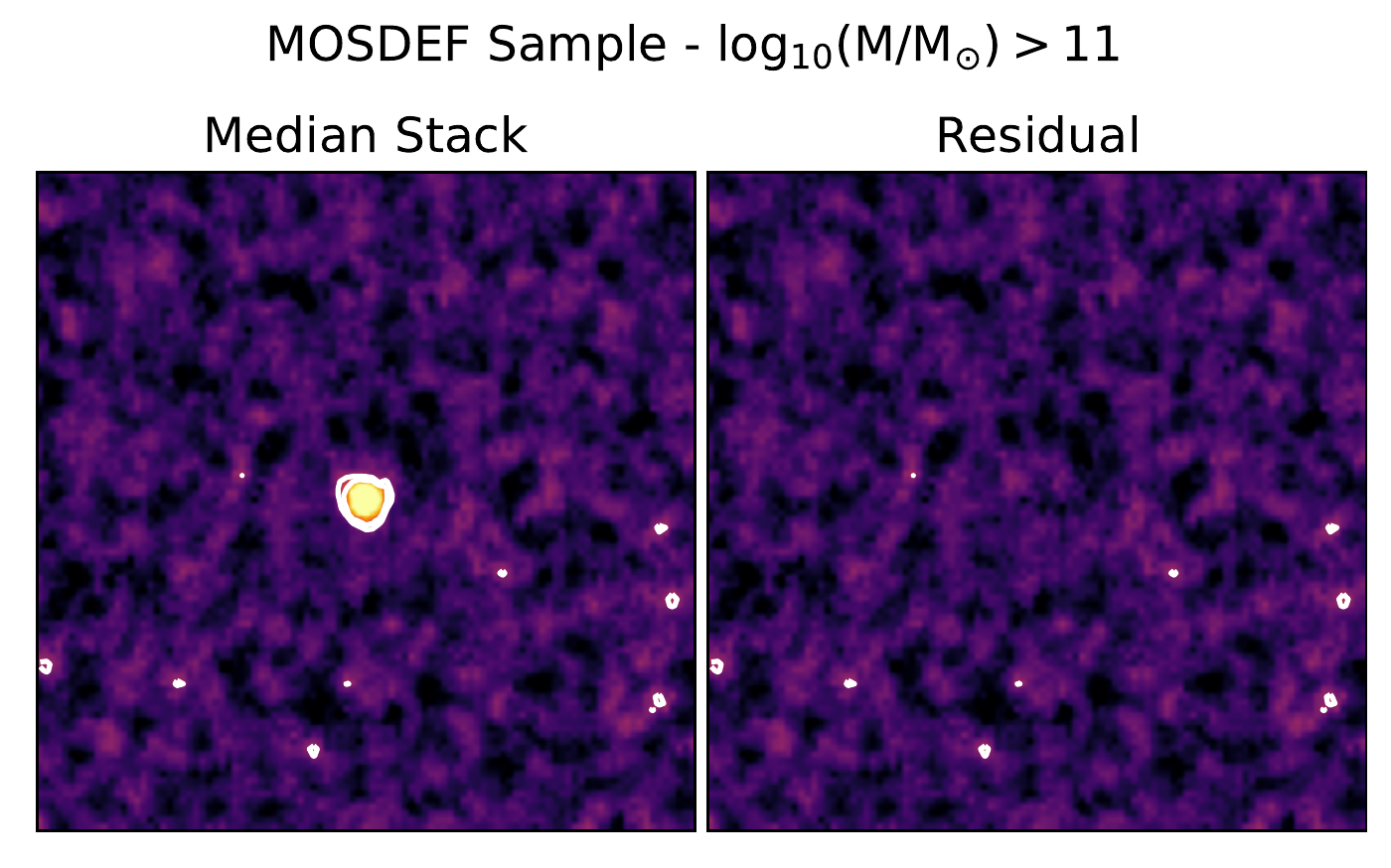}
  \caption{Left: Median stack of the 1.5 GHz VLA imaging for all sources in the MOSDEF sample with $\log_{10}(\text{M/M}_{\odot}) > 11$ and not classified as AGN (18 sources). Right: Residual image after subtraction of a 2D Gaussian beam to the median stack. We find no significant residual flux indicative of extended emission. In both images the colour scale stretches linearly from -2$\sigma$ to 15$\sigma$, where $\sigma$ is estimated from the robust scatter of the median stacked image. White contours represent 3 and 5$\sigma$ above the background noise.}
 \label{fig:m11_residual}
\end{figure}

Finally, we have assumed that sources are unresolved ($S_{\text{1.4GHz}}^{\text{Int}} =S_{\text{1.4GHz}}^{\text{Peak}}$) when measuring forced and stacked radio flux measurements.
This assumption is well motivated given the properties of the radio data in this analysis \citep{2020MNRAS.495.1188M}.
However, if the integrated-to-peak flux ratio for our stacked radio measurements is underestimated, we expect it to affect the most massive (and hence largest) galaxies.
In Fig.~\ref{fig:m11_residual} we show the 2D median stacked image for all $\log_{10}(\text{M/M}_{\odot}) > 11$ galaxies within the MOSDEF sample that are not classified as AGN (regardless of their H$\alpha$ S/N).
To test for extended emission we fit the stacked emission with a 2D Gaussian beam with the position and width fixed ($\text{FWHM} = 1.6\arcsec$).
After subtracting the model beam we find no statistically significant flux around the source within the residual map, with no evidence for extended emission.
Furthermore, we estimate a model-independent peak to integrated flux correction by measuring the total flux within a 5\arcsec\, diameter aperture, converting from Jy\,beam$^{-1}$ to Jy assuming the same 2D Gaussian beam.
Comparing the integrated flux to the peak flux for the stack we find $S_{\text{Int}}/S_{\text{Peak}} = 1.1\pm0.1$ for the $\log_{10}(\text{M/M}_{\odot}) > 11$ sample.
As noted in Section~\ref{sec:radio-data}, for the brighter radio population where integrated flux measurements have been made we found a median ratio of $S_{\text{Int}}/S_{\text{Peak}} = 1.07$ for the unresolved source population -- in excellent agreement with the estimate measured for the high mass stack.
Given these tests, we are confident that the systematic uncertainty in the radio luminosity measurements is of order $\sim10\%$.
As is the case for the other potential systematic uncertainties discussed above, the uncertainties on the radio luminosity measurements are not large enough to affect our key conclusions. 

\section{Discussion and Future Prospects}\label{sec:discussion}
With the next generation of radio continuum surveys, such as the LOFAR Two-meter Sky Survey (LoTSS) Deep Fields (Tasse et al. in prep, Sabater et al. in prep) 
and MIGHTEE \citep{2017arXiv170901901J},
we are now reaching sensitivities comparable to those probed by the VLA data in this study, i.e. typical SFRs for massive galaxies at $1 < z < 3$, but over 10s of deg$^{2}$.
Stacking analysis in these datasets using optically selected galaxy samples therefore has the potential to allow obscuration free SFR constraints out to the highest redshifts ($z > 5$) or in the lowest mass galaxies ($<10^{9}\,\text{M}_{\odot}$).
A key question for studies that aim to exploit this sensitivity is whether SFRs inferred from RC luminosities at higher redshifts are reliable.
In Section~\ref{sec:method} we have presented observations of the relation between dust-corrected H$\alpha$ and 1.4\,GHz radio luminosities for our MOSDEF sample in GOODS North.
From these results we conclude that for samples where the effect of AGN activity has either been removed or is not present, there is no evidence for redshift evolution in the SFR-RC out to $1.4 < z < 2.6$.

However, the caveat on identifying and/or excluding the contribution from AGN is extremely important.
In lower mass galaxies where luminous AGN activity is extremely rare, constraints on SFR from radio luminosity measurements (e.g. through stacking) should be reliable.
But for individual sources with the highest stellar masses where radio AGN activity may be prevalent, robustly identifying AGN within the sample remains a key step before radio luminosities can be used to measure obscured star-formation.
While we still do not fully understand the accretion history of SMBH out to $z < 2.6$, the rapid rise in radio AGN activity at the highest stellar masses has been observed from $z\sim0$ \citep{2019A&A...622A..17S} out to $z\sim1.2$ \citep{2011A&A...525A.127T}.
Furthermore, the stellar mass dependence of radio AGN activity is significantly steeper than for AGN activity identified through optical/X-ray criteria at a given redshift \citep{2011A&A...525A.127T}.
Given the combined optical/near-IR and radio sensitivities of multi-wavelength datasets used for studies of the redshift evolution of the FIRRC out to $z\sim3$ \citep[e.g.][]{2017MNRAS.469.3468C,2017A&A...602A...4D}, measurements based on individual sources are dominated by very massive galaxies at $z > 1$ - potentially leading to biases from unidentified low-level AGN activity \citep[see also][]{2018MNRAS.475..827M}.

The lack of significant redshift evolution in SFR-RC relation measured in this study supports those studies which infer no intrinsic evolution in the FIRRC \citep{2010ApJ...714L.190S, 2011MNRAS.410.1155B, 2010MNRAS.409...92J, 2014MNRAS.445.2232S}.
However, to fully understand the interplay between SFR, FIR and RC luminosities as a function of redshift, further studies are required.
With LoTSS Deep Field and MIGHTEE RC observations being located in fields with deep mid- to far-infrared observations required for robust measurements of the infrared luminosity and AGN/SF galaxy separation, robust constraints on the slope and evolution of the FIRRC over a broad range of redshift and stellar masses can now be made.

Furthermore, the forthcoming WEAVE-LOFAR spectroscopic survey \citep{Smith:2016vw} will also allow for a significant advance in our understanding of the SFR-RC and FIRRC at low redshift.
WEAVE-LOFAR will provide over $10^{6}$ high-resolution ($R\sim5000$) optical spectra for radio continuum selected galaxies across multiple survey tiers.
Together these tiers will allow for highly complete and representative samples of star-forming galaxies at $z < 0.4$ with robust measurements of dust-corrected H$\alpha$ SFR\footnote{We note that the $z < 0.4$ limit represents the regime where WEAVE can measure H$\alpha$ and H$\beta$ emission line luminosities. The full WEAVE-LOFAR sample will provide highly complete spectroscopic samples of radio continuum sources across a much higher redshift range - whereby the precise redshift information will allow for both improved radio source classification and physical modelling.}, enabling detailed statistical studies of the SFR-RC and its correlation with a wide range of physical properties - providing a clear reference point to which the results of this study can be compared.

\section{Summary}\label{sec:summary}
Sensitive radio continuum observations offer the potential for obscuration free measurements of star formation at higher resolution and over wider survey areas than currently possible with far-infrared surveys.
However, motivated by evidence for the apparent evolution in the far-infrared--radio correlation, questions remain regarding the redshift evolution of the star-formation rate - radio continuum (SFR-RC) relation.
In this study we test for redshift evolution in the SFR-RC relation out to $1.4 < z < 2.6$ using extremely deep 1.5 GHz radio continuum observations \citep{2018ApJS..235...34O} combined with dust-corrected H$\alpha$ SFR from the MOSFIRE Deep Evolution Field survey \citep[MOSDEF;][]{2015ApJS..218...15K}.

Thanks to the extensive multi-wavelength data available in the field we are able to reliably exclude AGN from our sample (through X-ray, mid-infrared and optical AGN diagnostics). 
Using stacking analysis to probe below the noise limits of the radio continuum imaging, we first measure the average radio luminosity for bins of dust-corrected H$\alpha$ luminosity in two redshift ranges. 
Next, to ensure our measurements are not biased by the selection criteria for dust-corrected H$\alpha$ measurements on individual sources, we perform a second analysis stacking both the rest-frame optical spectroscopy and radio continuum measurements as a function of stellar mass.
In both analyses we find that the measured radio luminosities are consistent with those predicted based on their H$\alpha$ derived SFR using $z\sim0$ relations from the literature.
Accordingly, we find no evidence for redshift evolution.

Given the available measurements, we conclude that SFR-RC relations measured at $z\sim0$ remain valid out to $z \lesssim 2.6$ within the typical intrinsic scatter measured for these relations ($\sim0.2$ dex).
However, the disagreements on the slope and normalization of the existing $z\sim 0$ relations present in the literature originate from uncertainties at the high SFR end of the SFR-RC relation at $z\sim 0$ ($\text{SFR} \gtrsim 20\, \text{M}_{\odot}\,\text{yr}^{-1}$).
Future spectroscopic and multi-wavelength photometric studies have the potential to improve measurements of SFR-RC at high SFR across a broad range of redshift and galaxy parameter space - further enhancing the potential for radio continuum luminosity as a robust SFR indicator. 

\section*{Acknowledgements}
The authors thank Frazer Owen for providing access to the VLA imaging used in this analysis as well as advice on its use.
KJD would also like to thank Philip Best and Daniel Smith for valuable discussions throughout the writing of the paper.
KJD acknowledges support from the ERC Advanced Investigator programme NewClusters 321271.
IS is supported by NASA through the NASA Hubble Fellowship grant \# HST-HF2-51420, awarded by the Space Telescope Science Institute, which is operated by the Association of Universities for Research in Astronomy, Inc., for NASA, under contract NAS5-26555.
Funding for the MOSDEF survey was provided by NSF AAG grants AST-1312780, 1312547, 1312764, and 1313171 and archival grant AR-13907, provided by NASA through a grant from the Space Telescope Science Institute.
The data presented herein were obtained at the W. M. Keck Observatory, which is operated as a scientific partnership among the California Institute of Technology, the University of California and the National Aeronautics and Space Administration. The Observatory was made possible by the generous financial support of the W. M. Keck Foundation. The authors wish to recognise and acknowledge the very significant cultural role and reverence that the summit of Maunakea has always had within the indigenous Hawaiian community. We are most fortunate to have the opportunity to conduct observations from this mountain.

\section*{Data Availability}
The spectroscopic data underlying this article will be shared on reasonable request to the MOSDEF co-PIs: \href{http://mosdef.astro.berkeley.edu/team/}{http://mosdef.astro.berkeley.edu/team/}.
Additional radio data underlying this article were provided by Frazer Owen by permission. Data will be shared on request to the corresponding author with permission of Frazer Owen.
The derived data generated in this research will be shared on request to the corresponding author. 





\bibliographystyle{mnras}
\bibliography{bibtex_library}







\bsp	
\label{lastpage}
\end{document}